\documentclass{aa}    
\usepackage{graphics} 
\def\HI{H{\sc i}\,} 
\begin{document} 
 
\thesaurus{ 
		11.09.1 The Cartwheel; 
		11.09.2; 
		11.19.3 
             } 
\title{Molecular Gas in The Cartwheel Galaxy
\thanks{based on observations made with the Swedish-ESO Submillimeter
Telescope (SEST) at La Silla, Chile}} 
   
\author{C. Horellou
	\inst{1},
	V. Charmandaris 
	\inst{2},  
	F. Combes
	\inst{2}, 
	P.N. Appleton
	\inst{3},
	F. Casoli
	\inst{2}
	\and	
	I.F. Mirabel  
	\inst{4,5}
	} 
\offprints{C. Horellou} 
  
\institute{
	Onsala Space Observatory, Chalmers University of Technology,
	S-43992 Onsala, Sweden
	\and	 
	Observatoire de Paris, DEMIRM, 61 Avenue de l'Observatoire, 
	F-75014 Paris, France 
	\and 
	Department of Physics \& Astronomy, Iowa State University, 
	Ames IA, 50011, U.S.A. 
	\and 
	Service d'Astrophysique, CEA-Saclay, F-91191 Gif sur 
	Yvette Cedex, France 
	\and
	Instituto de Astronom{\'\i}a y F{\'\i}sica del Espacio, cc 67, suc 28, 
	(1428) Buenos Aires, Argentina.
	} 
\date{Received 28 July; accepted 21 October 1998} 
\authorrunning{Horellou et al.} 
\titlerunning{Molecular Gas in The Cartwheel Galaxy}
\maketitle 
 
\begin{abstract}   
We present the first detection of molecular gas in the Cartwheel, 
the prototype of a collisional ring galaxy formed in the
head-on encounter of two galaxies. Until now, only very
little atomic gas and no CO had been detected in the centre, 
where gas is theoretically expected to pile up. 
Using the Swedish ESO Submm Telescope, we detected both
$^{12}$CO(1--0) and (2--1) line emission towards the central 
position. 
The line ratio 
and the line widths 
suggest that the CO(2--1) emission is sub-thermal
and that the CO(1--0) emission arises within the central 22$''$ 
(13 kpc);
it is probably associated with the inner ring and nucleus. 
We infer a mass of molecular gas (H$_2$) of 
1.5 to 6 $10^{9}$ M$_{\odot}$, 
which is significantly higher than the $\sim 10^8$ M$_\odot$ of atomic 
gas within that region. 
The low excitation of the gas, whether it is due to a low temperature or 
a low density,  is consistent with the weak star-forming activity observed
in the centre of the Cartwheel. 
\end{abstract} 
  
\section{Introduction} 

The Cartwheel Galaxy, discovered by Zwicky (\cite{zwicky}), earned its
nickname from the prominent system of spokes which connects the bright
outer ring to the inner ring (see Figure~\ref{co}). 
The nature of the Cartwheel was elucidated by early
numerical simulations, which showed convincingly that such ring systems
can form when a companion galaxy plunges through the centre of a
larger rotating disk (Lynds \& Toomre \cite{lt}; Theys \& Spiegel
\cite{ts}).  The passage of the
``intruder'' triggers a ring-wave that propagates through the disk of the
``target'' galaxy and induces massive star formation in a well-defined
ring.

In contrast to most merging galaxies in which star formation is
significantly enhanced in a very small area (the very centre), ring
galaxies exhibit starbursts on a large scale.  
Most observed ring galaxies have blue colours and elevated CO,
H$\alpha$, far-infrared and radio continuum emission (see Appleton \&
Struck-Marcell \cite{rings} for a review).  Because of the simple
geometry of the interaction, they appear as textbook examples to study
the effects that a collision has on the interstellar medium of a galaxy
and to follow the chronology of the star-formation in the expanding ring.

The Cartwheel Galaxy, 
because of its well-defined inner and outer rings, 
is considered as the prototype
of its class. It has been the subject of several observational
studies as well as dynamical modeling (Struck-Marcell \& Higdon
\cite{curt_jim}; Hernquist \& Weil \cite{hw};
Mihos \& Hernquist \cite{mihos}).  Optical and near-IR
imaging showed strong radial colour gradients in the disk interior to the
outer ring, which may trace the evolution of the stellar population in
the wake of the density wave (Marcum et al. \cite{marcum}).
HST images have revealed with unprecedented detail the
distribution of massive young clusters in the outer ring, as well as
the diffuse and knotty structure of the spokes (Borne et
al. \cite{borne}; Appleton et al. ~\cite{iau}).

The outer ring of the Cartwheel ($\sim$ 70$\arcsec$ in diameter) is
expanding, has blue colours and is populated by massive star-forming
regions (Higdon \cite{jim_ha}, Amram et al. \cite{amram}).  Most of
the star formation occurs 
in the southern quadrant of the ring, 
where the H${\alpha}$
emission, as well as the 20 cm and 6 cm radio continuum emission 
peak 
(Higdon \cite{jim_hi}).  Spectroscopy of a few
\ion{H}{ii} regions in the outer ring indicates a low metallicity
(Fosbury \& Hawarden \cite{fh}).  Most of the atomic gas (\HI) is
associated with the outer ring.
The
detection of a massive \HI plume extending more than 80 kpc in
projection in the direction of a companion galaxy (G3) suggests that G3
rather than one of the two nearer companions 
collided with the Cartwheel and produced its present-day appearance
(Higdon \cite{jim_hi}).

The inner ring ($\sim$ 18$\arcsec$ in diameter) and nucleus of the
Cartwheel seem gas-poor and until recently, no evidence of star
formation could be found. This is in disagreement with the models that
predict an infall of gas towards the centre and vigorous star
formation in the inner ring (Struck-Marcell \& Appleton \cite{cp}).
However, a number of recent observations has revealed a richer 
environment in the inner regions of the Cartwheel:

i) The HST images resolve a network of dust lanes in the inner ring
and luminous kiloparsec-size cometary structures which are suggestive 
of massive dense clouds traveling through the ambient gas 
(Struck et al. \cite{struck}).

ii) H$\alpha$ emission has been detected at a low level 
throughout the inner ring and nucleus (Amram et al. \cite{amram},
Higdon et al. \cite{jim_baas}). 

iii) ISOCAM images show strong 7\,$\mu$m and 15\,$\mu$m emission
in the inner ring and nucleus. The mid-IR fluxes  
indicate the presence of hot dust and are consistent with weak
star formation activity (Charmandaris et al. \cite{vassilis}).

These new results strengthened our expectation that molecular gas must
be present in the central region of the Cartwheel.
Previous searches for $^{12}$CO(J=1--0) line emission 
had been unsuccessful, and the Cartwheel was one of the two
galaxies that remained undetected in the survey of 16 ring galaxies of
Horellou et al. (1995).  
Here, we present
the first $^{12}$CO(1--0) and $^{12}$CO(2--1) detections towards the
centre of the Cartwheel. For $H_0$=75\,km\,s$^{-1}$\,Mpc$^{-1}$,
the distance to the galaxy is 121 Mpc.

  
\section{Observations and Data Reduction}   

We have observed at $\alpha$=00h37m40.8s, $\delta$=$-33^{\circ}$42$'$56.9$''$ 
(J2000), 
which is the peak of emission of the Cartwheel nucleus in the HST
I--band image.

The observations have been carried out in July 1998 in La Silla
(Chile) with the 15m Swedish-ESO Submillimeter Telescope (SEST) (Booth
et al. \cite{booth}).  We used the IRAM 115 and 230 GHz receivers to
observe simultaneously at the frequencies of the $^{12}$CO(1--0) and
the $^{12}$CO(2--1) lines.  At 115 GHz and 230 GHz, the telescope
half-power beam widths are 43$''$ and 22$''$, respectively.  The main-beam
efficiency of SEST is $\eta$$_{\rm mb}=T_{\rm A}^*/T_{\rm mb}$=0.68 at 115 GHz and
0.46 at 230 GHz (SEST handbook, ESO).  
The typical system temperature varied between 220 and 320 K 
(in $T_{\rm A}^*$ unit)
at both frequencies. The total on-source integration time was 6 hours. 
A balanced on-off dual beam
switching mode was used, with a frequency of 6 Hz and two symmetric
reference positions offset by 12$'$ in azimuth.  The pointing was
regularly checked on the SiO maser R Aqr.  
The pointing accuracy was 4$''$ rms. 
The backends were
low-resolution acousto-optical spectrometers.  The total bandwidth
available was 500 MHz at 115 GHz and 1 GHz at 230 GHz, with a velocity
resolution of 1.8 km s$^{-1}$.  The data were reduced with the
software CLASS. 
Only first-order baselines were subtracted
from the spectra.   

\section{Results and Discussion}

\begin{figure*} 
\resizebox{18cm}{!}{\includegraphics{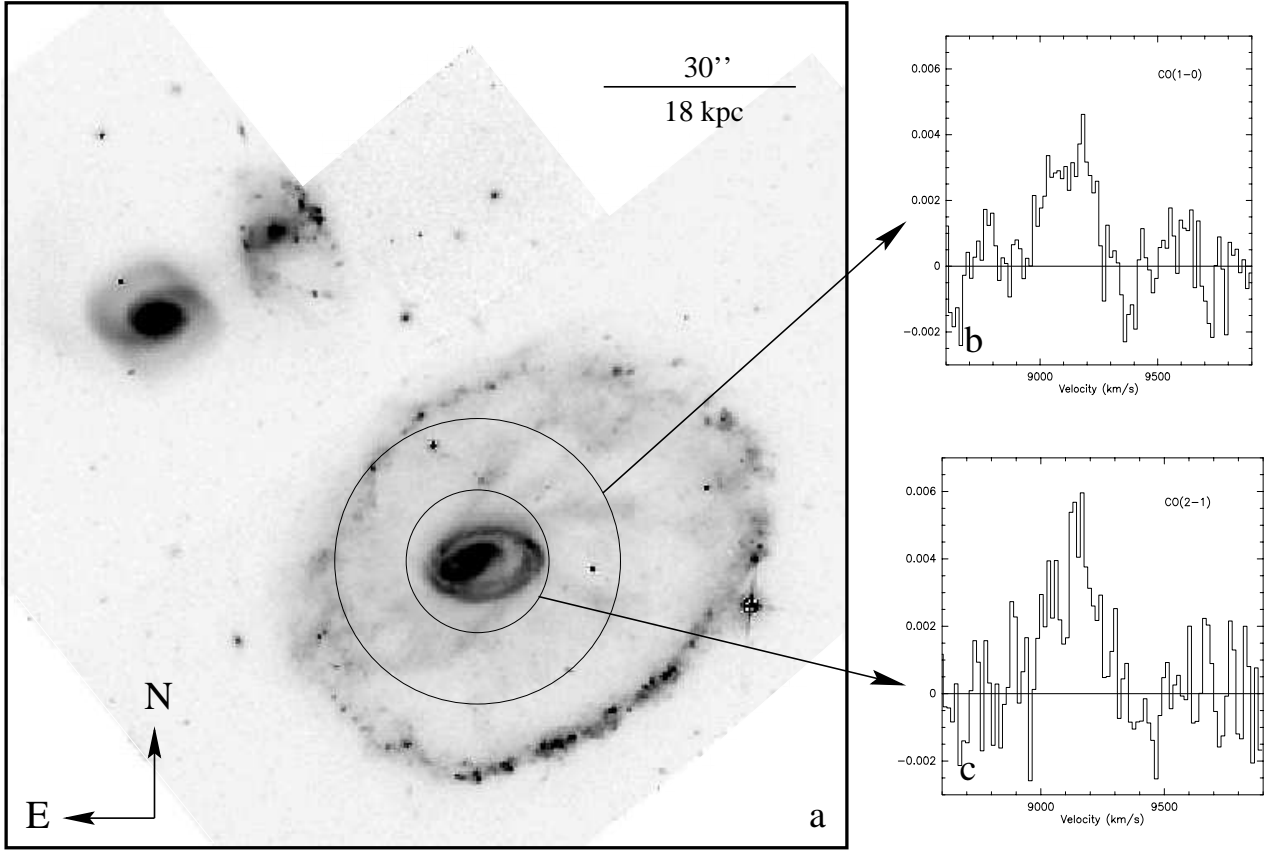}}  
 \caption{
a) HST I--band image of the Cartwheel galaxy (Borne et al. \cite{borne_hst}). 
The circles on the optical image represent the resolution
of the CO(1--0) and CO(2--1) observations (43$''$ and 22$''$
respectively).
b) The CO(1--0) spectrum of the nuclear regions. 
The $y$-axis shows the main-beam temperature in K.  
The velocity resolution is 15 km s$^{-1}$. 
c) The CO(2--1) spectrum. Units and velocity resolution are the same as
in b).
}
  \label{co} 
\end{figure*}

In Figure~\ref{co} we present the CO(1--0) and CO(2--1) spectra
obtained towards the centre of the Cartwheel, as well as the HST image
(Borne et al. \cite{borne_hst}) 
on which the sizes of the CO beams have been superimposed. The parameters of
the lines, derived from gaussian fits 
after smoothing to a final velocity resolution of 15 km s$^{-1}$, 
are given in Table~\ref{results}. 
The central  
velocities agree with the systemic velocity of 9089\,km\,s$^{-1}$
determined from \ion{H}{i} observations (Higdon \cite{jim_hi}). 
An indicative molecular gas mass within the CO(1--0) beam 
is given in the last column of Table~\ref{results}.
It has been computed using a standard CO-H$_2$
conversion factor (see below). 

\begin{table}[h]  
\caption[ ]{Observational  results.} 
\label{results} 
\begin{flushleft}  
\begin{tabular}{llllll}  \hline
Line	& $v_{\rm hel,opt}$	& ${\Delta}v$ & $\int{T_{\rm mb}dv}$  
&$\sigma_{\rm mb}$  	& M(H$_2$)\\
 	& km s$^{-1}$	& km s$^{-1}$ & K km s$^{-1}$ 	  & mK 		&10$^9$M$_{\odot}$\\
\hline
CO(1--0) & 9123$\pm$13	& 218$\pm$23      & 0.82$\pm$0.10   & 1.0	&1.5\\
CO(2--1) & 9136$\pm$12	& 211$\pm$27      & 0.95$\pm$0.11   & 1.2	&\\
\hline    
\end{tabular} 
\end{flushleft}   
\end{table}

\subsection{Molecular gas mass and CO(2--1)/(1--0) line ratio}

To convert CO intensities into H$_2$ column densities,
we use the factor of Strong et al. (\cite{strong}):

\begin{equation}
$$X= N(H_2)/I_{\rm mb}(CO)= 2.3\, 10^{20} {\, \rm mol\,cm^{-2}\, (K km\, s^{-1})^{-1}} $$
\end{equation}

\noindent where $I_{\rm mb}(CO)= \int T_{\rm mb}dv$ is the main-beam line area.
This yields a mass  

\begin{equation}
$$M(H_2)\, ({\rm M_{\odot})}
\,=\, 
1.25\,10^5\,(\theta/43'')^2\,I_{\rm mb}(CO)\,(D/1{\rm Mpc})^2  $$
\end{equation}

\noindent where $\theta$ is the half-power beam width of the telescope. 

From the CO(1--0) line intensity, we estimate a mass of molecular gas
(H$_2$) of 1.5 10$^9$ M$_{\odot}$ within the 43$''$ beam. 
This value should be interpreted with caution since it is known that
the CO-to-H$_2$ conversion factor ($X$) is dependent on the metallicity
(e.g. Maloney \& Black \cite{maloney}).  
Due to the very weak line emission in the centre, 
the metallicity could be 
measured only in the outer star-forming ring of the Cartwheel, and it is 
12+log[O/H]=8.1 (Fosbury \& Hawarden \cite{fh}),
compared with 8.9 in Orion. 
If the nucleus has the same metallicity as the outer ring, then the
conversion factor would be higher than the standard one 
by a factor of about 4 
(Wilson \cite{wilson}), and the H$_2$ mass as well.
However, it is also known that most galactic disks present a  
metallicity gradient, with metallicities in the nucleus up to $\sim10$
times higher than in the outer parts of the disk (Smartt \& Rolleston
\cite{smartt}).
Although $X$ may vary linearly with the reciprocal of metallicity 
when the metallicity is low, $X$ is expected to become less sensitive 
to metallicity at values above that of Orion 
(Sakamoto \cite{sakamoto}). 
Thus, if we take into account the metallicity gradient across the disk,
the value of $X$ for the nucleus of the Cartwheel might be not much
lower than the standard value. 
We can bracket the conversion factor in the nucleus: 
$X < X_{\rm nucl} < 4 X$, 
and the H$_2$ mass  
within the 43$''$ beam: 
$1.5 \,  10^9\,  < {\rm M(H}_2{\rm)\, [M}_\odot] < 6 \, 10^9. $

H$_2$ masses in this range are common 
for central regions of normal spirals of
intermediate and late types (e.g., see Fig. 133 in Young et al. 
\cite{young}) 
and are similar to those inferred for other ring galaxies
(Horellou et al. \cite{horellou}). 

More unusual is the fact that we observe similar integrated intensity
in the CO(1--0) and CO(2--1) lines, despite the 
difference of a factor of 4 
in the beam areas.  
If the CO lines were thermalized, 
this would imply that the surface density of the CO gas 
would be nearly constant within the 25 kpc CO(1--0) beam diameter, 
which is very improbable. 
The surface density of most disk galaxies 
shows an exponential radial decrease. 
Even if the spokes were gas-rich, as some numerical simulations predict
(Hernquist \& Weil \cite{hw}, Mihos \& Hernquist \cite{mihos}),
their filling factor is likely to be low.  
For a linear decrease of the surface density, one expects a line ratio of 0.5. 
We rather suggest that 
the CO(2--1) emission is sub-thermal, possibly because of a low temperature 
or a low gas density, 
and that the CO emission is concentrated within the 13 kpc CO(2--1) beam 
(which is consistent with the fact that both lines have similar widths). 
Correcting for the different beam sizes, one obtains a 
CO(2--1)/CO(1--0) line ratio of 0.3 
$\pm 0.05$,
which is significantly lower 
than the ratio of 0.9 observed in nearby spirals (Braine \& Combes
\cite{braine}). Such low ratios have been observed in dark clouds of M31 
where there is little evidence of star-forming activity 
(Allen \& Lequeux \cite{allen}).

\subsection{Comparison of CO, \HI and 
H$\alpha$ emission}

\begin{figure} 
\resizebox{\hsize}{!}{\includegraphics{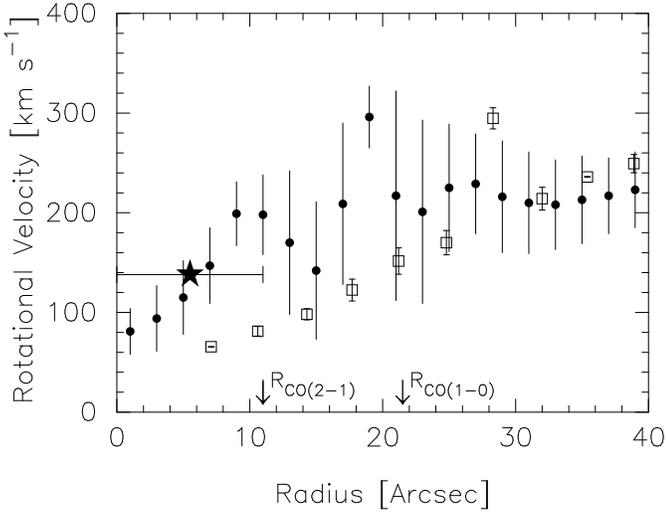}}  
 \caption{
The rotation curve of the Cartwheel assuming an inclination of 50$^\circ$. 
The open squares are the \ion{H}{i} values (Higdon \cite{jim_hi}), 
the filled circles are the H$\alpha$ points (Amram et al. \cite{amram}) and the 
filled star is the velocity derived from the CO(2--1) line width. 
The arrows on the $x$-axis indicate the radii of the beams of the CO(1--0)
and CO(2--1) observations. 
}
  \label{vrot} 
\end{figure} 

Let us try to compare the spatial distribution and kinematical
signature of the CO gas with those of the atomic and the ionized
hydrogen.  
Neutral hydrogen is 
undetected within the central $R=10\arcsec$
of the galaxy ($\Sigma$(\HI) $<$ 0.3 M$_\odot$ pc$^{-2}$)
and has a roughly constant surface density from $R=10''$ to 22$''$. 
($\Sigma$(\HI) $\approx$ 4 M$_\odot$ pc$^{-2}$, 
Higdon \cite{jim_hi}). 
The \HI mass within our 43$''$ CO(1--0) beam
is 1.5 10$^9$ M$_\odot$, which is of the same order as the molecular gas
mass that we have calculated, but it is only  $\sim 10^8$ M$_\odot$
within the 22$''$ CO(2--1) beam. If, as argued in the previous paragraph, 
the CO emission is concentrated, then the interstellar medium within the
central 22$''$ is predominantly molecular.

The $\ion{H}{i}$
rotation curve rises smoothly from the inner to the outer ring 
(see Figure~\ref{vrot}). 
Atomic gas is infalling near the
inner ring, where $\ion{H}{i}$ concentrations are seen.
Assuming that the CO gas is distributed in an 
inclined plane 
($i\sim50^\circ$), we can use the line widths 
to estimate the rotational velocity 
in the central region of the galaxy. 
For the CO(2--1) line we find 
$v_{R21}$=${\Delta}v/{2{\rm sin} i}$= 138 km s$^{-1}$
which is significantly higher than the velocities measured
for the \HI gas at radii  
$R < 20 \arcsec$ (see Fig.~\ref{vrot}).
The CO point is more consistent with the H${\alpha}$ rotation curve. 
The difference between the \HI velocity on one hand 
and the H$\alpha$ and CO velocities on the other hand is likely to be due to 
different distributions of the atomic and molecular gas, the latter being more
concentrated.
 
\section{Conclusion} 
We have presented the first detection of $^{12}$CO(1--0) and $^{12}$CO(2--1) 
line emission towards the nucleus of the Cartwheel galaxy
and estimated the mass of molecular gas 
within the central 43$''$ (1.5 10$^9 <$ M(H$_2$) $< 6\, 10^9$ M$_\odot$). 
The limited angular resolution of these single-dish
observations does not allow us 
to determine precisely the location of the molecular gas, 
but it is probably  
concentrated within the central $22''$ where little atomic gas is found
($\sim 10^8$ M$_\odot$, Higdon \cite{jim_hi}).
The CO(2--1) to (1--0) line ratio is low, suggesting sub-thermal excitation, which 
is consistent with the low level of star-forming activity observed in the central 
region of the Cartwheel. 
These CO measurements make it possible to justify  
future interferometric observations
to study the distribution
and the dynamics of the molecular gas with a better spatial resolution 
and to establish whether  
it is associated with the inner ring and/or the dust-lanes, 
and whether it is of pre-collisional origin or infalling onto the nucleus 
as numerical simulations predict. 

\begin{acknowledgements}
We are grateful to L.-{\AA}. Nyman and to the SEST staff for the support during
the observations, to S. Leon (Observatoire de Paris), 
C. Struck (Iowa State University)  
and J.H. Black (Onsala Space Observatory) 
for useful discussions and comments.
C.H. aknowledges financial support from the Swedish Natural Science Research
Council (NFR). 
V.C. acknowledges the financial support from the TMR fellowship
grant ERBFMBICT960967. We would like to thank an anonymous referee for critical 
comments on the manuscript.  
\end{acknowledgements} 

\end{document}